\documentclass[twocolumn]{aa}

\usepackage{graphicx}

\usepackage{txfonts}

\begin{document}

\title{High-speed, energy-resolved STJ observations of the AM Her
system V2301 Oph}

\author{A.P. Reynolds, \inst{1}
           G. Ramsay, \inst{2}
           J.H.J. de Bruijne, \inst{1}
           M.A.C. Perryman, \inst{1}
           M. Cropper,\inst{2}
           C.M. Bridge, \inst{2}
           A.Peacock \inst{1}
          }

   \offprints{M. A. C. Perryman: mperryma@rssd.esa.int}

   \institute{$^{1}$Research and Scientific Support Department of ESA, ESTEC,
              Postbus 299, 2200 AG Noordwijk, The Netherlands\\
              $^{2}$Mullard Space Science Laboratory, University College London, 
              Holmbury St Mary, Dorking, Surrey RH5 6NT, UK
             }

\date{Accepted; Received; in original form }

\authorrunning{Reynolds et al}

\abstract{We present high time-resolution optical energy-resolved
photometry of the eclipsing cataclysmic variable V2301~Oph made using
the ESA S-Cam detector, an array of photon counting super-conducting
tunnel junction (STJ) devices with intrinsic energy resolution. Three
eclipses were observed, revealing considerable variation in the
eclipse shape, particularly during ingress. The eclipse shape is shown
to be understood in terms of AM~Her accretion via a bright stream,
with very little contribution from the white dwarf photosphere and/or
hotspot. About two thirds of the eclipsed light arises in the
threading region.  Variation in the extent of the threading region can
account for most of the variations observed between cycles.  Spectral
fits to the data reveal a 10,000\,K blackbody continuum with strong,
time-varying emission lines of hydrogen and helium. This is the first
time that stellar emission lines have been detected in the optical
band using a non-dispersive photon-counting system.

}

\maketitle

\section{Introduction}

V2301~Oph is an interacting binary system with an orbital period of
113 min (Silber et al.\ 1994). The spectropolarimetric study of
Ferrario et al. (1995), made during a low accretion state, established
that its accreting white dwarf has a magnetic field strength of
$B\sim7$~MG. On this basis, V2301~Oph is a polar (a magnetic
cataclysmic variable, CV), with the weakest magnetic field of any
member of this class. In these systems, matter flows from the donor
star along a ballistic stream, until it reaches a `threading region'
where it becomes attached to the magnetic field of the white
dwarf. The material then travels along the field lines, until it forms
a column of material settling onto the surface of the white dwarf via
a shock (e.g. Ferrario \& Wehrse \ 1999). In some systems (eg HU~Aqr,
Bridge et al.\ 2002) the details of the accretion geometry can be
reliably related to the shape of the eclipse lightcurve. The highly
variable nature of the eclipse profiles in V2301~Oph, however, make
the interpretation of the lightcurve more challenging.

With its unique combination of high time-resolution and energy
discrimination, S-Cam2 -- ESA's energy-sensitive photon-counting
camera -- has proven particularly useful as a tool for studying the
rapidly varying optical emission of CVs. Technical descriptions of the
instrument may be found in Rando et al. (1998, 2000), while the data
acquisition and reduction methods are outlined in Perryman et
al. (1999, 2001). In our previous studies of CVs (Perryman et al.\
2001, Bridge et al.\ 2002, 2003, Steeghs et al.\ 2003) our main tool
for understanding the spectral changes has been energy-selected
lightcurves and their associated colour ratios. Here, we present
S-Cam2 data of V2301 Oph and relate the light curves to previous
observations and demonstrate that S-Cam2 can detect and quantify
contributions from individual emission lines of hydrogen and helium.

\begin{table}
\begin{flushleft}  
\begin{tabular}{ccccc}
\hline
No & Date & Cycle & Obs & Phase Range \\
    & (2000) & & (UTC) & \\
\hline
1  & April 27 & 45769---45770 & 03:52--04:28 & 0.802--1.115\\
2  & April 28 & 45781---45782 & 02:17--03:03 & 0.706--1.106\\
3  & April 29 & 45793---45794 & 01:03--01:43 & 0.800--1.154\\
\hline
\end{tabular}\\
\end{flushleft}
\caption{Observations of V2301~Oph obtained using S-Cam2. The
ephemeris of Barwig et al.(1994) has been used to determine the
binary orbital phases.}
\label{tab:observations}
\end{table}

\section{Observations}

The S-Cam2 campaigns were performed during April 2000 at the 4.2-m
William Herschel Telescope on La Palma, with the instrument mounted at
the Nasmyth focus: the seeing was variable, ranging from good to
poor. The observations were planned so as to include three
complete eclipses, with representative data either side. Due to
constraints from other program targets, it was not possible to observe
an entire cycle.  The observations are logged in Table~1. We estimate
that V2301 Oph was at $V\sim$15.7, indicating it was in a high
accretion state.

\begin{figure*} 
\centering
\includegraphics[angle=-90,width=15cm]{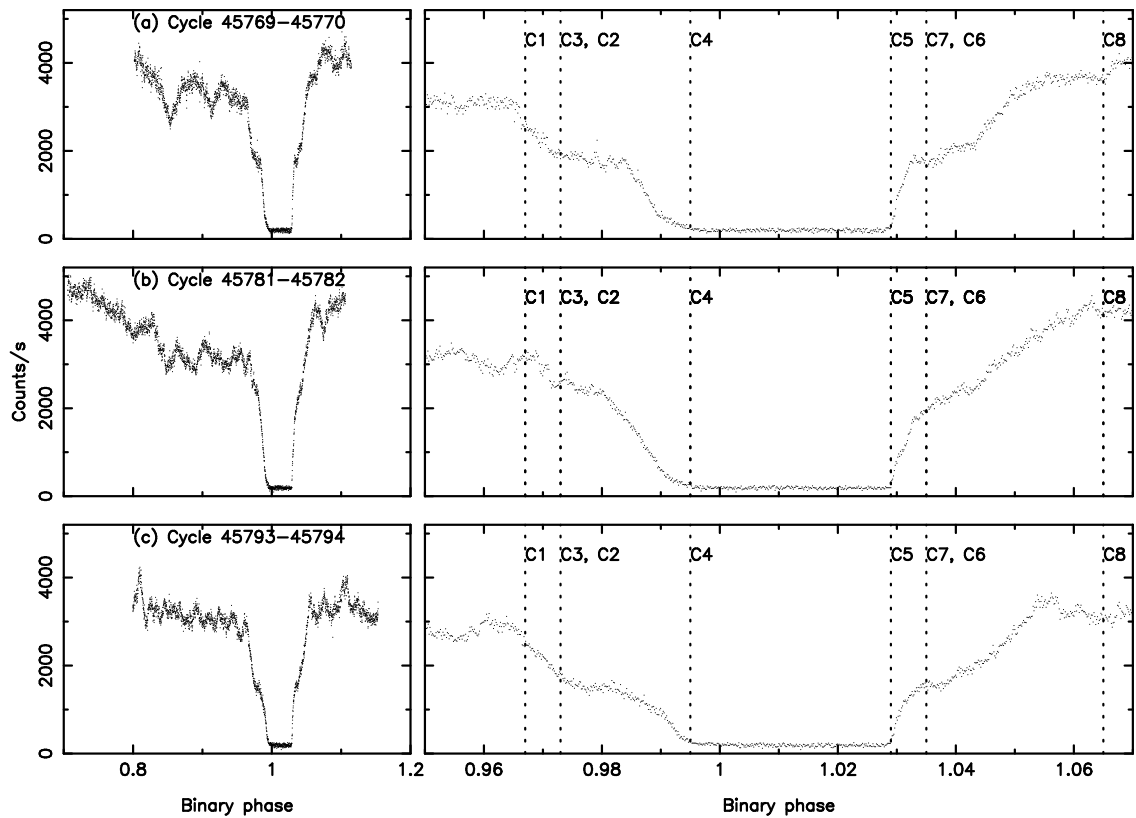}
\caption{Phase-folded light curves of the three V2301~Oph
observations.  The right-hand plots are enlargements of the eclipse
intervals in the left-hand plots.  These white-light ($\lambda
\sim3200-7200$ \AA), background-subtracted light curves have been
corrected for variations in pixel-to-pixel gain and responsivity. A
time (i.e., zenith angle) dependent correction for atmospheric
extinction has also been applied. The data have been binned into 1~s
intervals.  The phases and cycle numbers are with respect to the
ephemeris of Barwig et al.\ (1994). The vertical dashed lines indicate
the phases of the contact points identified by Steiman-Cameron \&
Imamura (1999); c$_{1}$ and c$_{5}$ are the contact points of
photospheric ingress and egress respectively.}
\label{Figeclipses} 
\end{figure*}

In Fig \ref{Figeclipses} we show `white-light' lightcurves for the
three observations, after barycentric correction, background
subtraction, phase-folding and correction for atmospheric extinction;
see Perryman et al.\ (2001) for a description of these data reduction
steps. The data have also been corrected for variations in
pixel-to-pixel detection efficiency and energy response. The light
curves show that the accretion stream in V2301 Oph is particularly
prominent in the optical band, compared with many other polars.

\section{Light Curves}

Our data were phased with respect to the orbital ephemeris of Barwig
et al.\ (1994), which was also the ephemeris used by Steiman-Cameron
\& Imamura (1999). There are significant variations between the 3
eclipse profiles in each of the eclipses. They do, however, exhibit a
similar shape to those seen in other studies. The first is similar to
the 1997 June 2 eclipse seen by Steiman-Cameron \& Imamura (1999),
with a pronounced two-stepped descent into the eclipse. This
two-stepped profile is also seen in the UV data of Schmidt \& Stockman
(2001).

Steiman-Cameron \& Imamura (1999) identify eight contact points in the
lightcurve, associated with the ingress/egress of the white dwarf
photosphere and the ingress/egress of the accretion stream. By
reference to Fig~6 in Steiman-Cameron \& Imamura, we determined the
phases of their contact points and then plotted these phase points as
vertical dashed lines in our Fig~1. They associate c$_{1}$, c$_{2}$,
c$_{5}$ and c$_{6}$ with the eclipse of the photosphere, while they
associate the four other contact points with the eclipse of the
accretion stream. We show the observed times of these contact points
in Table \ref{contacts}.

The contact points as defined by Steiman-Cameron \& Imamura (1999)
generally match up with the same feature in our light curves. For
contact points c$_{4,5}$ the features line up very well. There is some
discrepancy in the eclipse ingress because the accretion stream in
V2301 Oph is relatively bright compared to many other polars. In the
second eclipse, the stream accounts for roughly 3/4 of the optical
emission. We believe that discrepancies between the other contact
points can be explained by a variation in the amount of material
located in different regions of the accretion stream trajectory. We
discuss this in greater depth in \S 5.

\begin{table}
\begin{center}
\begin{tabular}{lrrrr}
\hline
Contact & Phase &   Eclipse 1 &   Eclipse 2 &   Eclipse 3 \\
Points &        &      &  & \\
\hline
C1  &  (0.967)  & 51661.17442  &   51662.11582 &  51663.05722\\
C2,C3 & (0.973) &  51661.17489 &    51662.11629 &  51663.05769\\
C4   & (0.995) &  51661.17661 &    51662.11802 &  51663.05942\\
C5   & (1.029) &  51661.17929 &    51662.12069  & 51663.06209\\
C6,C7 & (1.035) &  51661.17975  &   51662.12116 &  51663.06255\\
C8  &  (1.065) &  51661.18211 &    51662.12350 &  51663.06491\\
\hline
\end{tabular}
\end{center}
\caption{The times for the contact points for the three eclipses. The
times are in MJD. The error on each time estimate is $\sim$3 sec.}
\label{contacts}
\end{table}

\section{Spectral properties}

The intrinsic energy sensitivity of the S-Cam2 detector permits high
time-resolution spectrophotometry of astronomical sources.  The
resulting data can be analysed in terms of energy (colour) bands and
their associated ratios (Perryman et al.\ 2001) or even direct
spectral fitting (de Bruijne et al.\ 2002, Reynolds et al.\ 2003).  In
this paper we extend this spectral-fitting to accommodate the presence
of emission lines, a particularly useful technique for cataclysmic
variables due to the swiftness of the eclipse transitions. First,
however, we examine the data in terms of energy bands, since this is
useful for characterising broad changes in emission properties.

\subsection{Colour bands}

The data were segregated into three bands corresponding to blue (3200
--- 4700 \AA), visual (4700 --- 5500 \AA) and red (5500 --- 7200
\AA\hspace{1mm} colours). We label these bands B$^{\prime}$,
V$^{\prime}$ and R$^{\prime}$, emphasising that they are not on any
formal photometric system.  The boundaries between the bands are
selected to ensure that roughly equal numbers of events are assigned
to each band.  In Fig~2, we again show the three eclipses, but this
time split into these three energy-selected lightcurves. The
amplitudes of the V$^{\prime}$ and B$^{\prime}$ band curves are
adjusted slightly to correct for small differences in total counts
(the initial segregation only giving approximate equipartition among
events) and the V$^{\prime}$ and B$^{\prime}$ bands are then offset
vertically on the same plots.  Differences in slope between the bands
are therefore meaningful.  Although there are no dramatic differences
between the bands, in all three eclipses the R$^{\prime}$ band data
exhibits noticeably more structure, with the steps and the ledges
between them more sharply defined than in the other bands. In
particular, the first and third eclipses both contain dips in the
R$^{\prime}$ band at phase $\sim$ 0.0375, which are much less apparent
in the B$^{\prime}$ and V$^{\prime}$ data. In the first eclipse, as
well, the ledge between the two ingress steps is more or less
horizontal, whereas it shows a pronounced slope in the B$^{\prime}$
and V$^{\prime}$ data. This suggests that if the material responsible
for the steps is concentrated in two distinct regions, then the region
between them is emitting mainly at shorter wavelengths than that seen
in the R$^{\prime}$ band data, and/or is preferentially absorbed in
the red.

The relatively small numbers of events per time bin in each colour
band make the study of colour ratios difficult. In order to remove
short-term variations due to seeing, guiding errors and intrinsic
flickering, we first smoothed the individual R$^{\prime}$ and
B$^{\prime}$ band lightcurves using a sliding cell with a width of
15\,s. The ratio R$^{\prime}$/B$^{\prime}$ against binary phase, for
all three eclipses, is shown in the narrower panels in Fig~2. It is
apparent that there are significant differences in the variation of
R$^{\prime}$/B$^{\prime}$ between the three observations, particularly
in the late stages of the ingress, just before the onset of deep
eclipse. 

\begin{figure}
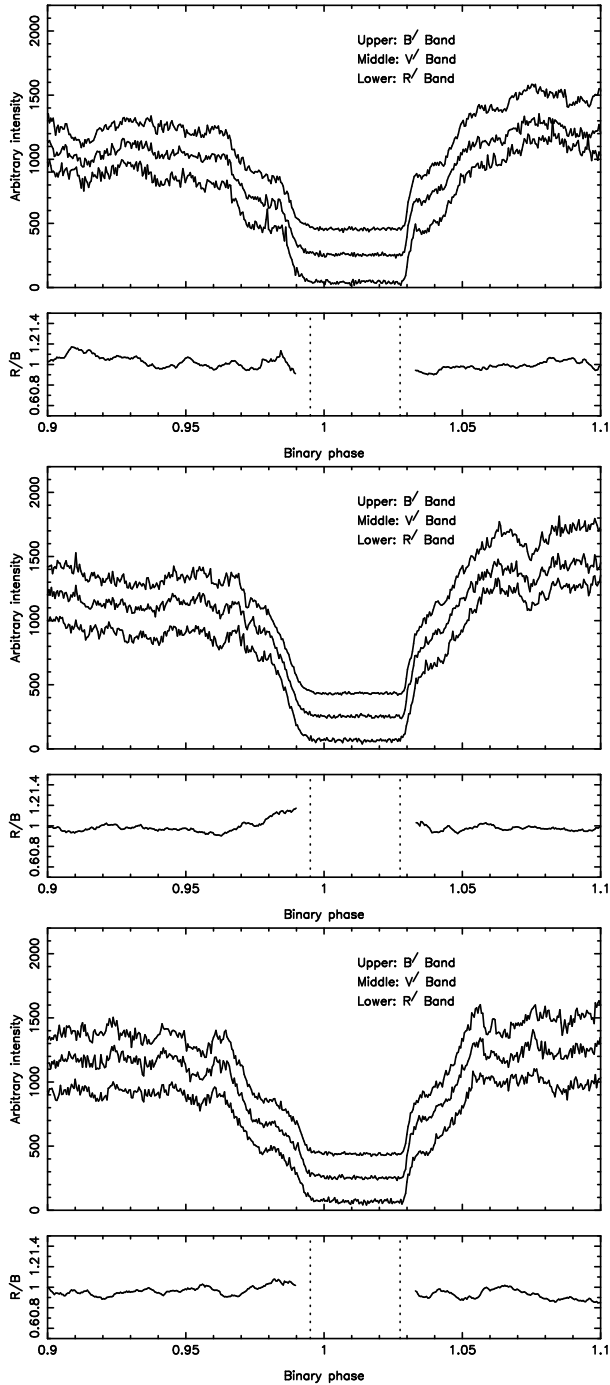

\centering
\includegraphics[angle=-90,width=8cm]{hardness_1.ps}
\includegraphics[angle=-90,width=8cm]{hardness_2.ps}
\includegraphics[angle=-90,width=8cm]{hardness_3.ps}
\caption{Energy-band selected lightcurves for the three eclipses in
Fig~1, with the ratio R$^{\prime}$/B$^{\prime}$ against phase plotted
below each observation. The vertical dotted lines in the
R$^{\prime}$/B$^{\prime}$ plots indicate the approximate start and end
phases of the deep eclipse.  Not only is the eclipse ingress and
egress shape in general more sharply defined in the R$^{\prime}$ band
data than the B$^{\prime}$ band, but there are significant differences
between the three ratio plots.}
\label{Figsimlcs}
\end{figure}

\subsection{Spectral fitting}

We now examine the same data using the complimentary approach of model
fitting to time-resolved spectra.  Each complete eclipse observation
was sliced into consecutive 25~s intervals, with a single spectrum
created from each interval. The duration of the intervals was chosen
as a compromise between time-resolution and S/N, and is short enough
to probe spectral variations on the typical timescale of the major
ingress/egress features seen in the lightcurves.  The spectra were
corrected for electronic artifacts and pixel-to-pixel gain variations.
Corresponding background spectra were obtained by using 150~s data
intervals from the deepest part of the eclipse in all three
observations, ensuring that any residual contribution from the donor
star is removed from the spectrum, in addition to the sky.

Spectral fitting made use of the XSPEC model-fitting package (Arnaud
1996), developed for use in high-energy astrophysics, but adaptable to
the analysis of low-resolving power optical data such as S-Cam
spectra.  In the left-hand panels of Fig~3, we show the result of
fitting a blackbody model against a post-eclipse slice (the final 25~s
interval) from the first observation, together with the associated
residuals. The optical flux therefore originates from the white dwarf,
the accretion region and the accretion stream. The reduced $\chi^{2}$
for this fit was 4.8. The systematic structure in the residuals
clearly indicates the inadequacy of this single-component model.

In the right-hand panels of Fig~3, we show the improvement to the fit
by the addition of emission lines to the blackbody model. We fixed the
blackbody temperature to be 10000 K which was the mean value derived
after experimenting with $kT_{bb}$ as a free parameter (the range in
$kT_{bb}$ was a few 1000K). The lines included in our fit were
H$\alpha$ and H$\beta$, both of which are seen in emission in the data
of Barwig et al.\ (1994), Silber et al.\ (1994), Ferrario et al.\
(1995) and Simic et al.\ (1998), and He\,{\sc i} 5875, observed
strongly in emission in many AM~Her spectra, including V2301~Oph
(Barwig et al.\ 1994). He\,{\sc ii} 4686 is also likely to be a strong
feature of the spectrum. This line, however, lies sufficiently close
to H$\beta$ that, with our limited resolution, the two lines are
blended.  While our fit statistics show that it is necessary to have
an emission component in this part of the spectrum, we are more than
likely modelling the contribution of two lines with our single
component labeled 'H$\beta$'.

Using the published optical data of Simic et al., we estimated that
the width ($\sigma$) of H$\alpha$ is $\sim$ 170 \AA. The line widths
in our model were therefore fixed at this value, and the centroid
wavelengths were fixed at their laboratory values.  We also added a
'Broad Emission Component' centered on 3540 \AA. This is to take into
account the many features expected to exist to the blue side of
He\,{\sc ii} 4686, including the Balmer limit, which it would be
impractical to fit as individual components given our limited
resolution. The width of this component was allowed to vary, but for
the other lines the only adjustable parameter was the
normalisation. With all model parameters present, the mean reduced
$\chi^{2}$ over the first eclipse observation was 1.67, compared to
6.0 for the blackbody alone.

\begin{figure}
\centering
\includegraphics[angle=-90,width=8cm]{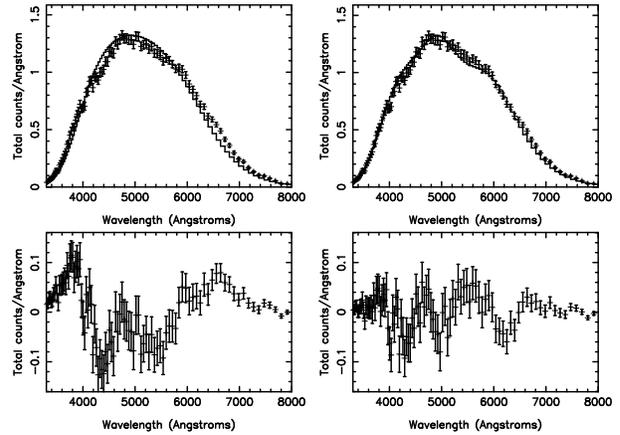}
\caption{Left: blackbody fit, with associated residuals,
for the final 25\,s slice in the first observation. Right: the
same data, but now modelled with a blackbody and emission lines.
The residuals are plotted on the same scale as in the left-hand
box.}
\label{Figsimlcs}
\end{figure}

\begin{figure} 
\centering
\includegraphics[angle=-90,width=8cm]{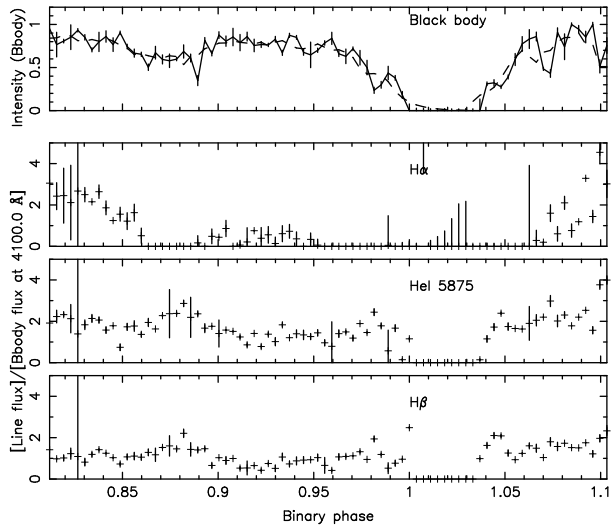}
\caption{Continuum and line strengths in observation 1. From top to
bottom: phase-resolved, normalised line intensities for the following
model components: blackbody (for T$_{\rm eff}$ = 10\,000~K),
H$\alpha$, He\,{\sc i} 5875, and H$\beta$. The line strengths are
normalised to the instantaneous blackbody intensity, and refer to the
height of the line as a fraction of the continuum height at 4130 \AA.
The final model line, the Broad Emission Component, is not plotted,
since its identification is uncertain and it varied erratically with
time.}
\label{Figsimlcs1}
\end{figure}

In Figure \ref{Figsimlcs1} we show the variations in line strengths
with binary phase for the first eclipse observations. The mean
$\chi$$^{2}$ for the second and third observations is worse than for
the first: increasing to 1.9 and 2.4 respectively. This is probably
due to long-term drifts in the gain, after warming and cooling of the
cryogenic system.

The top panel of Figure \ref{Figsimlcs1} shows the variation in the
normalisation of the blackbody, together with a smoothed version of
the same trend. It is seen that this curve closely resembles the shape
of the white light lightcurve: this is expected since most of the
photons detected will come from the continuum rather than the lines.

The trends in the three narrow lines in the model are plotted below
the blackbody normalisation on the same time axis. In order to make
the line strengths meaningful, we normalised them to the blackbody
continuum strength: the plotted intensities represent the height of
the line above the continuum, as a fraction of the continuum height at
4130 \AA.  If the line intensities are proportional to the underlying
blackbody normalisation therefore, the resultant trend will be
approximately flat. This is seen to be the case for He\,{\sc i} 5875
and H$\beta$. H$\alpha$, however, becomes progressively fainter
relative to the above lines, almost vanishing around $\phi$=0.95. It is
likely that H$\beta$ also becomes weaker, but that this is masked by
the contribution from He\,{\sc ii} 4686 (see above). One physical
reason for the Balmer lines to become weaker as the phase nears
eclipse is that they are more likely to originate from the heated face
of the secondary star (and hence not visible at eclipse), while the
He\,{\sc ii} 4686 line originates from the accretion stream, which is
visible for longer. This also explains why H$\alpha$ is also somewhat
slow to reappear after the sharp rise in continuum intensity during
the egress, displaying a $\sim0.02$ lag in phase. The final model
parameter, the Broad Emission Component at 3540 \AA\hspace{1mm} varies
erratically, but is generally stronger in the post-eclipse phases than
prior to the eclipse.

\section{Discussion}

In \S 3 we showed that the optical emission from V2301 Oph has a
larger contribution from the accretion stream compared to many other
polars.  In order to explore this further, we trace the Roche
potential out of the binary system along the line of sight from any
point within the vicinity of the white dwarf.  We then adjust the
system parameters, (mass ratio (q) and inclination (i), and accretion
spot location), under the constraints of grazing contact of this line
of sight with the Roche lobe at particular phases. We initially
adopted (q,i) values of Steiman-Cameron \& Imamura (1999), namely 0.25
and $81.75^{\circ}$ respectively. However, with the usual assumption
that the stream follows a ballistic and then a magnetic trajectory, we
found that the stream would be fully eclipsed by $\phi$=0.99, whereas
our observations show that the stream remains visible until phase
$\phi\sim$0.995. In order to resolve this, the donor star must become
smaller, with a corresponding increase in the inclination. There are
no individual constraints on (q,i), only the pairs set by the eclipse
length, as illustrated by the curve in Fig~7. The choice of q=0.15,
i=$84.0^{\circ}$ is sufficient to ensure stream visibility at phase
0.99, so in the absence of further constraints we base our subsequent
discussion on these values.

\begin{figure}
\centering \includegraphics[angle=-90,width=8cm]{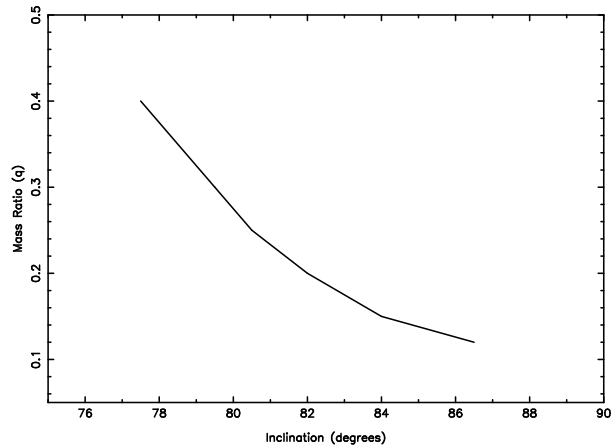}
\caption{The relationship between mass ratio (q) and inclination for
V2301~Oph, as defined by the eclipse geometry. The phase of the stream
visibility favours q=0.15, i=$84.0^{\circ}$.}  
\label{Figeclipse}
\end{figure}

\begin{figure}
\begin{center}
\setlength{\unitlength}{1cm}
\begin{picture}(8,16)
\put(-1.4,0.8){\includegraphics{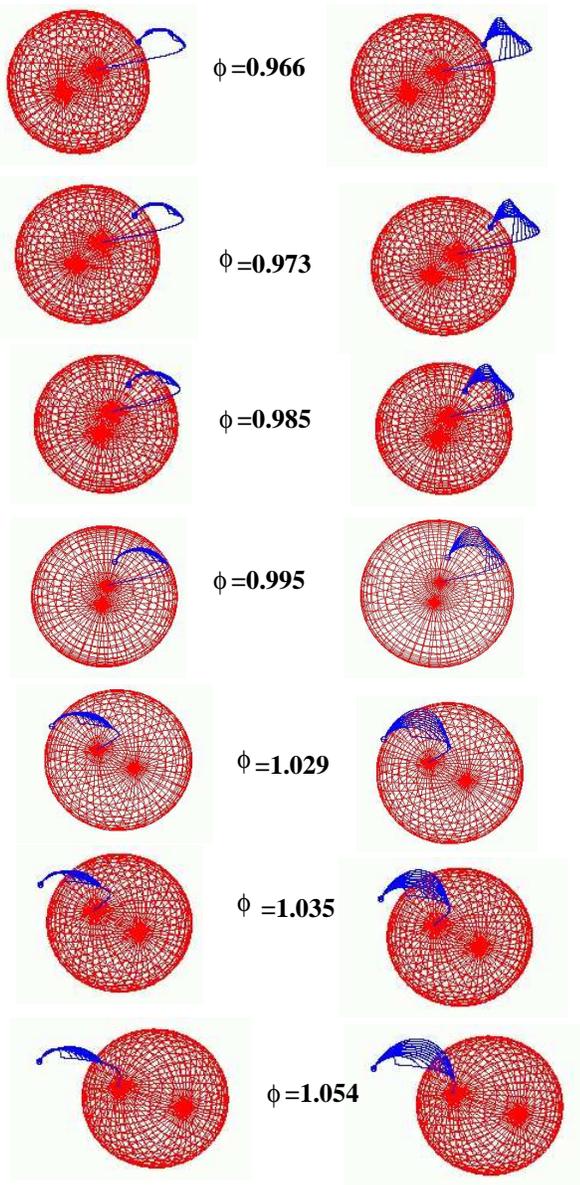}}
\end{picture}
\end{center}
\caption{The eclipse geometry of V2301~Oph, for q=0.15,
i=$84^{\circ}$. The spot azimuth is $60^{\circ}$, and the magnetic
field is aligned at $20^{\circ}$ to the white dwarf spin axis. The
plots on the left show six phases of the eclipse for the case where
the threading occurs later in the stream, while the plots on the right
show the same six phases for the early threading scenario.}
\label{binary_plot}
\end{figure}

In Figure \ref{binary_plot}, we show the orientation of the binary at
different eclipse phases, for two different configurations of the
threading region. The parameters used to generate these plots, apart
from (q,i), are a spot azimuth of $65^{\circ}$ and the magnetic field
aligned at $20^{\circ}$ to the white dwarf spin axis, both of which
are in line with the values in Steiman-Cameron \& Imamura (1999). The
threading of material onto the magnetic field lines takes place
between 0.2 --- 0.12 $a$ (where $a$ is the binary separation) for the
set of plots on the left, and between 0.4 --- 0.12 $a$ for the set of
plots on the right.

The scenario on the left can then be related to the eclipse lightcurve
seen in the first observation (Fig~1). By $\phi$=0.967, the white
dwarf is fully eclipsed. This causes only a small dip in the
lightcurve, however, and is only really apparent in the first
observation. By $\phi$=0.973, the bright part of the magnetic stream
closest to the white dwarf has also been eclipsed. The intensity then
declines only gradually until $\phi$=0.985, when the threading region
begins to be occulted. Brightness declines rapidly, and by
$\phi$=0.995 the stream is entirely eclipsed.

On egress, the first component to emerge is the white dwarf,
accounting for the sudden, steep rise in intensity. This is followed
immediately by the hot stream, characterised by a slightly shallower
slope. There is then another plateau as the empty mid-arch of the
magnetic field is revealed. At $\phi$=1.035 the threading region again
begins to emerge, and by $\phi$=1.054 the threading region is
completely visible. The implication of this observation is that around
one third of the stream brightness is in the region around the white
dwarf, while the remaining two thirds arises in the threading region.

Turning to the scenario on the right, an increase in the extent of the
threading region can explain the eclipse shape of the second
observation. At $\phi$=0.973, the threading region is already starting
to be eclipsed, and by $\phi$=0.985 is substantially hidden. However,
the mid-arch of the magnetic field contains more material and is
visible for longer. This accounts for the longer, brighter decline to
eclipse seen in the second observation. As before the stream is fully
eclipsed by $\phi$=0.995. At $\phi$=1.054, during the egress, the
threading region is still not completely revealed, and hence the
lightcurve continues to increase in brightness beyond this phase, in
contrast to the first observation.

We therefore have a self-consistent explanation for the white-light
variations seen in the first two observations, within the context
of the standard ballistic-plus-magnetic stream. Almost all of the
observed light arises in the stream, and two thirds of that is
contributed by the threading region. Observation three closely
resembles observation one, except that the threading region appears
somewhat less bright. 

The colour information, as presented in Section~4.1, also supports the
above interpretation. Fig~2 shows that until the stream becomes rather
faint at $\phi$=0.99, the second observation is redder than the
first. This shows that what we are seeing in the second observation is
cooler material in the furthest reaches of the threading region,
whereas in the first observation, we are seeing material plunging deep
into the field before threading. Observation three (in which there was
a slight reddening compared to the first observation) is a little
harder to understand, but the threading region is fainter here anyway,
so is perhaps smaller and cooler than in the first observation.

\section{Conclusions}

We have presented energy-resolved STJ photometry of the eclipsing
polar V2301~Oph, reporting the first detection and analysis of stellar
emission lines based on non-spectrally dispersed optical data. Three
eclipses are observed in total, with significant variations in shape
observed between them. These variations are shown to be consistent
with the usual AM~Her accretion scenario, whereby mass flows from the
donor star to the white dwarf along a ballistic stream, before
becoming threaded onto the magnetic field lines of the white dwarf.
By varying the extent of the threading region along the ballistic
stream, the major features of the observations can be accounted for,
provided the mass ratio is decreased to 0.15, with a corresponding
increase in the inclination to i=$84.0^{\circ}$. The colour variations
seen in the energy-resolved data are also consistent with this
interpretation, since they suggest that the material is at its hottest
when it penetrates most deeply into the magnetic field before becoming
threaded.

These observations show that current STJ detectors can resolve optical
emission lines. However, the resolution of such detectors is much
poorer than can be obtained with dispersive spectrometers.  Until now,
nearly all STJ work has used the materials niobium and tantalum. The
obvious next step is to migrate the technology to even lower
critical-temperature super-conductors such as hafnium ($T_{\rm c} =
0.128$~K, compared to $4.48$~K for tantalum). With a band-gap well
below a meV (0.021~meV), hafnium has the potential to provide better
energy resolution at all energies and is therefore of keen interest,
for both visible and X-ray applications. The predicted tunnel-limited
resolving power of hafnium is $\sim$64 at 1~eV, so that the wavelength
resolution is $\sim$2 \AA\hspace{1mm} at Ly~$\alpha$, $\sim$50
\AA\hspace{1mm} at 5000 \AA, and 150 \AA\hspace{1mm} at 20000 \AA.

\begin{acknowledgements}

We thank the ING staff at the WHT on La Palma for their excellent
technical support during all the S-Cam2 telescope campaigns. The
other members of the optical STJ development team at ESTEC are
also thanked for their work on the S-Cam program.

\end{acknowledgements}

\end{document}